\documentstyle{kluwer}
\begin{opening}
\title{Geometrical parametrization of warps for edge-on galaxies}
\author{
J. \surname{Jim\'enez-Vicente},
 C. \surname{Porcel},
 M. L. \surname{S\'anchez-Saavedra}, 
 E. \surname{Battaner}}
\institute{Depto. de F\'{\i}sica Te\'orica y del Cosmos. Univ. de Granada.
18071 Granada. Spain}
\date{}
\end{opening}

\begin{document}

\maketitle

\begin{abstract}

        We propose new parameters to describe the geometry of a warped disc
when viewed edge-on, a global warp parameter $w$, and a family of three 
parameters
$A$, $B$ and $C$ that describe, independently, the shape of the warp. 
These parameters are useful to detect some key effects in
the understanding of warps. We have also
developed software (WIG) which is able to calculate these parameters for a
given image of a galaxy in any wavelength and which is described here with
some examples, namely ESO 235-53 in the optical, NGC 4013 in 21 cm and NGC 4565
in 1.2 mm continuum.
\end{abstract}

\section{Introduction}

        The warp of most discs of spiral galaxies is at present a controversial
dynamic phenomenon. Current hypotheses have been reviewed by Binney (1992),
Combes (1994) and Battaner (1995). From the observational point of view a
statistical analysis is at present lacking. Some works have been reported
dealing with the statistics of warps, such as those by Bosma (1991), Briggs
(1990) and Christodoulou et al. (1993). But these works consider a small sample
of 21 cm mapped galaxies, mainly because few galaxies at present have available
21 cm maps. These galaxies have been studied in considerable detail, on the
other hand. Warps are worse observed in the optical, but the available sample
is bigger. Statistical analysis of optical warps has been carried out by
S\'anchez-Saavedra, Battaner and Florido (1990) and Reshetnikov (1995). However
there are at present large quantities of data, in particular the {\em Digitalized
Sky Survey}, which could allow these works to be greatly extended.

        With this purpose, it would be convenient to develop software to
automatically characterize the basic properties of warps, avoiding subjective
appreciations, and to define some parameters accounting for the basic
description of each warp. The number of these parameters should be kept small,
whilst retaining a geometrical description as complete as possible. We introduce
here a {\it warp parameter}, accounting for the degree of warping, and three
fitting parameters for the warp curve.

        We also describe here the software developed to obtain these parameters. 
This software, called WIG (Warps in Inclined Galaxies) precisely
determines the centre and the position angle of the galaxy, cleans the image
from nearby stars, calculates the centroid curve and determines the value of
the above-mentioned warp parameters.

        We show examples of the application of WIG to different
edge-on galaxies observed at different wavelengths. In the optical (I
band), we study the galaxy ESO 235-53, which exhibits a clear warp,
using the data from de Grijs (1997).The warp of this galaxy has been
studied in detail by de Grijs (1997). 
In 21 cm we examine the well known warp of NGC 4013 (one of the most 
representative warped spirals) using the data 
from Bottema (1995). We also study the galaxy NGC 4565 as it has been observed
in the mm continuum by Neininger and Guelin (1995), whose emission is mainly
produced by dust, and which represents a very interesting new tool to study
warps, as well as other dynamic features of galaxies.

\section{The warp parameters}

        The warp of a galactic disc is a global property of the galaxy, and
so, its measurement should not be affected by the internal details.
Therefore, it is convenient to have a simplification of the galactic geometry
which allows us to study global geometric properties. With this purpose in
mind we construct what we call the {\it warp curve}. When studying the warp 
for an edge-on galaxy, we are interested in two directions contained in the 
plane of the sky: the direction defined by the major axis of the galaxy and the
direction of the rotation axis of the galaxy. Throughout this paper we will 
identify
the former with the $x$ direction and the latter with the $y$ direction, taking
the centre of the galaxy as the origin of coordinates. Then, the {\it warp
curve} is defined as the locus of points $x_i$, $y_i$ which tells us the
deviation with respect to the symmetry plane of a point at a given distance from
the galactic centre. The warp curve represents just the warp when the galaxy is
projected in the plane of the sky, and cannot account for effects such as the
twist of the line of nodes. Bottema (1996) also suggests that a corrugated dust
lane can mimic the presence of a warp. However, it is not known whether an
external corrugation and a warp are different phenomena from a kinetic point of
view. There are also other, lesser, difficulties of interpretation. Nevertheless, 
the warp curve is an adequate
description of the geometry of the projected warp, and therefore, 
we will take it
as the starting point in the definition of the geometrical parameters.

        A galaxy is said to be more warped than another if the {\it deviation} 
of its {\it outer} part with respect to the plane of symmetry is greater. 
According to this concept, and taking into account the previously defined axis, 
a parameter which is intended to account for the warp should be proportional to 
the $y$ coordinate of all points in the warp curve. On the other hand, the warp 
is a peripheral phenomenon, and therefore, the outer points should have a 
greater weight when measuring the warp. A properly defined ``warp parameter''
should match the folloing properties:

        a) It should be non-dimensional, to assure its value does not depend on
        the chosen units (pixels, arcsec, etc...)
        
        b) It should not depend on the galaxy size (only on its shape)
        
        c) It should not depend on the angular resolution of the image (even if
        a better resolution would provide a more precise evaluation).
        
        The continuous form of the definition should be $\int y x dx$, or
        better, to get a non-dimensional expression, dividing each quantity by
        L (the galaxy size): $\int (x/L)(y/L)(dx/L)=(1/L^3)\int y x dx$. The
        discrete form must therefore be

\begin{equation}
w=\frac{\Delta}{L^3} \sum_i x_i y_i
\end{equation}

where $\Delta$ is the pixel size. This definition is valid for any chosen unit.
In particular, if we take the pixel as unity, so that $\Delta=1$, we define the
warp parameter as

\begin{equation}
w=\frac{1}{L^3} \sum_i x_i y_i
\end{equation}

        The absolute value of $w$ is a measurement of
the degree of warping, and the sign of $w$ distinguishes between N-like and
S-like warps.

        Sometimes the warp of a disc is not completely symmetric, and then
this parameter would hide the information of the clearly warped side of the
galaxy. To avoid this problem we also define the warp parameters on each side
of the galaxy independently. Therefore, the right warp parameter is defined as:
\begin{equation}
w_r=\frac{1}{4 L_r^3} \sum_{x_i \geq 0} x_i y_i
\end{equation}
measuring $x_i$ and $y_i$ in pixels. $L_r=max(x_i)$ is the size of the right 
side of the galaxy. Similarly, the left warp parameter is defined as:
\begin{equation}
w_l=\frac{1}{4 L_l^3} \sum_{x_i \leq 0} x_i y_i
\end{equation}
measuring $x_i$ and $y_i$ in pixels. $L_l=max(-x_i)$ is the size of the left side of the galaxy.

        With this parameter we can, for example, compare the warp of a galaxy
at several wavelengths to determine whether or not a colour gradient exists 
within the warp. We can also, with the help of this parameter, detect warps 
that would otherwise have been undetected.

        We also propose some parameters which account for these macroscopic 
geometrical
features of the warp. To do this we fit each side of the warp curve to the
function:
\begin{equation}
y=\left\{
\begin{array}{ll}
0       & |x|<|A| \\
C\left( |x-A|-B \left( 1-e^{-\frac{|x-A|}{B}}\right) \right) & |x|\geq|A|
\end{array}
\right.
\label{ajuste1}
\end{equation}
        This function reproduces the shape of a warp, i. e., it is flat up to a 
point and then deviates from the symmetry plane until it reaches an asymptotic
direction. The interpretation of the parameters
$A$, $B$ and $C$ is as follows:

        .- $A$ is the starting point of the warp. $A$ has dimensions of length.
        
        .- $B$ is the characteristic length in which the warp reaches the
        asymptotic direction. $B$ has dimensions of length.
        
        .- $C$ is the value of the asymptotic slope. $C$ is adimensional.
        
        It has been observed in some galaxies that the warp begins in a given
direction, and then turns back to the mean plane and ends in the opposite
hemisphere. Part of this effect may be due to the fact that the line of nodes
and the line of sight do not coincide. Or it may be due to the existence of a
more warped dust lane. Or it may be due to an intrinsic
effect and indicate a real property of warps. For such galaxies a
four-parametric fitting would have been better. But an excessive number of
fitting parameters makes the interpretation of simple warps unclear, and we
have preferred a three-parametric fitting.

\section{The software: WIG}

        We have 
developed specific software (WIG) to calculate the warp curve
and the previously defined parameters form the image of an edge-on galaxy. In
the rest of this section we briefly describe how the software works.

        WIG has been made to work with nearly centred galaxies in horizontal 
position (i. e. the major axis of the galaxy approximately coincides with the 
$x$ axis and the centre of the galaxy is the centre of the image) as input. 
Therefore any image must be preprocesed by any standard package in order to 
fulfill this requirement.

        The first step that WIG takes is the calculation of the centre 
$(x_0,y_0)$ and the size of the galaxy 
$(\sigma_x,\sigma_y)$ (the centre should already be close to the centre of the
image as stated before, but this step calculates it more precisely). To do this
we use two alternative methods. An iterative gaussian fitting is the choice
when the centre of the galaxy has the maximum emission (as happens with optical
images); otherwise, a mean value equally weighted for all the 
points with an emission over one standard deviation of the sky noise is the 
choice (we term this method {\em Homogeneous Signal Minus Noise} (HSMN)). 
(During this step the software also calculates an estimation of the size
of the galaxy $(\sigma_x,\sigma_y)$ which is the width of the gaussians in the 
first case and the standard deviations in the second one). Once this has been
achieved, we select the image zone within 
$([x_0-3\sigma_x, x_0+3\sigma_x],[y_0-4\sigma_y, y_0+4\sigma_y])$.

        The next step is to erase the stars in the selected zone.
This step is a necessary one, because foreground stars lying close to the
galaxy can be brighter than it (especially in the peripheral zone in which
we are particularly interested), and therefore can lead us to erroneous
results. An example of this effect is shown at the top of figure (\ref{st1}).

The method chosen for this step has been as follows:
        We scan each row in the selected zone and  by means of a gaussian
fit look for peaks exceeding 
at least two standard deviations the mean sky noise. If the FWHM of the peak is 
less than a limit value (which we fix at
5 pixels), the pixels around the peak (a FWHM on each side) are substituted by
the noise value in that zone. This method has two main advantages over the
traditional two-dimensional fitting: first, this method is better at preserving
vertical gradients in the zone close to the galaxy, and, furthermore, it is
computationally more efficient, because we have less parameters in the fit. 
The result of this process is shown at the bottom of
figure (\ref{st1}).

\begin{figure}
\vspace{10cm}
\caption[]{Images of PGC 29691 before (top) and after (bottom) the star deleting
process.
}
\label{st1}
\end{figure}

        The next step is the calculation of the warp curve. We need to
calculate the {\it centres} of the galaxy in the direction perpendicular to its
symmetry plane. This is the crucial step in the whole process, and therefore 
we should be especially cautious. Again, two alternative methods are proposed
for this purpose:

        .- Gaussian fit: We fit each column in the selected zone to a gaussian.
The peak position gives us the {\it centre} we are looking for, but the FWHM
will also be used afterwards.

        .- {\em Shorth}: We select, for each column, the shortest interval
containing a given percentage of the data (we have used a value of$40 \%$ 
but other values around $50 \%$ lead to equivalen results), and then
we calculate the mean position of the peak and its standard deviation
in this interval.

        Once this step is completed we have not yet finished, because many of 
the columns belong to the sky background, and not to the galaxy. We have, 
therefore, to
select, among all the columns, those belonging to the galaxy. To do this we
scan the columns from left to right and in the opposite direction, and choose
the columns fulfilling the following requirements:

        1) The peak exceeds by at least one standard deviation the sky noise.

        2) The FWHM of the peak is smaller than $3\sigma_y$.

        3) The change in the peak position from one column to the next is
        smaller than $\sigma_y$.
 
        4) The change in the FWHM of the peak from one column to the next is
        smaller than $\sigma_y$.

        In the left to right scan, once we have marked $\sigma_x$ or more 
columns as belonging to the galaxy, the first column which does not fulfill the
requirements marks the right side of the galaxy. The same procedure in the 
opposite direction marks the left side of the galaxy.

        At this point we almost have the warp curve, but another step is still
necessary. The reason for this is that we must be sure that the axes in the
warp curve are the right ones, i.e. that the $x$ axis coincides with the
symmetry plane of the inner disc and the $y$ axis with the spin axis of the
galaxy. This is a crucial point, because slight deviations from this situation
would lead us to erroneous results. To be sure this condition holds we fit
the inner part of the warp curve (the pixels in the interval
$[x_0-\frac{3}{4}\sigma_x, x_0+\frac{3}{4}\sigma_x]$) to a straight line, and
then we rotate the warp curve the angle indicated by the slope of this line.
Moreover, the value of the ordinate at the origin is taken as the new
coordinate of the centre ($y_0$). This is the last step in the process of
calculating the warp curve, and now we are ready to start the calculation of
the previously defined parameters. This calculation is straightforward from
their definition and therefore does not need any comment.

        Although in most cases WIG gives good results, there are a few
cases in which it has problems. These are the cases of low inclination galaxies
in which the spiral arms dominate the geometrical aspect, discs with a marked 
dust lane, and images in which a large object is very close to the galaxy under
study. 

The value of the lower inclination angle for which the method is valid is
an essential parameter. Though this number is difficult to estimate,
our experience shows that for angles over about $80^\circ$ the method
gives reliable results.

Even though the method is limited to very edge-on galaxies (e.g. inclinations
higher than $80^\circ$), this limitation still keeps a rather large
sample available which is enough for a statistical study. In any case
this new method constitutes a sensible improvement with respect to the
subjective method, which has already given good results for optical
data.

\section{WIG at work: Some examples}

        Three examples have been chosen to illustrate the usefulness of this 
software and the parameters: an image in the optical, an image in 21
cm and an image in 1.2 mm.

        The first example will be the galaxy ESO 235-53 using the recent
data from de Grijs (1997) in the I-band. In this case, several large
foreground stars should be masked by hand before using the image as
input for WIG. We use the shorth method to calculate the warp curve (which is shown in  
figure (\ref{opt1}) superimposed to a contour map of the
galaxy). Before calculating the parameters, the
final warp curve has some regions which were linearly interpolated to
avoid the effects of the masks for the stars. This will slightly affect the
final results specially for the right part of the galaxy.

\begin{figure}
\vspace{10cm}
\caption[]{Warp curve for galaxy ESO 235-53 superimposed to a contour
  map of the galaxy. The big circles inside the galaxy belong to the
areas which were masked out to avoid foreground stars.}
\label{opt1}
\end{figure}

        With this curve we calculate the warp parameters. The results are:
        
\begin{eqnarray*}
w&=&0.0054 \\
w_r&=&0.0043 \\
w_l&=&0.0069 \\
A_r&=&3.4 \; {\mathrm kpc}\\
B_r&=&16.4 \; {\mathrm kpc}\\
C_r&=&0.175 \\
A_l&=&-15.3 \; {\mathrm kpc}\\
B_l&=&2.4 \; {\mathrm kpc}\\
C_l&=&0.223
\end{eqnarray*}

        Now, we show the results of WIG for the galaxy NGC 4013 from the 21 cm
data obtained from Bottema (1995). This galaxy is an excellent one to study 
warps, because it has an inclination angle of $90^\circ$ and the line of nodes
coincides with the line of sight. The warp of this galaxy has been extensively 
studied in the optical by Florido et al. (1991). The first step is again to
prepare the galaxy to
be used as input for WIG (i.e. to put it in a horizontal position). Now we will
use the HSMN as the method for calculating the centre (because now the centre
of the galaxy is not the brightest part, and therefore a gaussian fit would not
be appropriate), and the gaussian fit for calculating the warp curve. The
resulting warp curve, superimposed onto the 21 cm image is shown in figure
(\ref{cm1})

\begin{figure}
\vspace{10cm}
\caption[]{Image of NGC 4013 in 21 cm 
with its warp curve superimposed.}
\label{cm1}
\end{figure}

        The warp parameters calculated for this case are:
        
\begin{eqnarray*}
w&=&-0.0228 \\
w_r&=&-0.0232 \\
w_l&=&-0.0225 \\
A_r&=&6.07 \; {\mathrm kpc}\\
B_r&=&3.20 \; {\mathrm kpc}\\
C_r&=&-1.144 \\
A_l&=&-7.90 \; {\mathrm kpc}\\
B_l&=&1.75 \; {\mathrm kpc}\\
C_l&=&-0.599
\end{eqnarray*}

        This galaxy is extraordinarily warped as can be seen from the value
of the parameter $w$, and its warp is very symmetric, as shown by the
similarity between the parameters $w_r$ and $w_l$. This galaxy is, moreover, a
singular one, because its warp in the optical and in 21 cm points in opposite
directions.

        Finally we show an example in the millimetre range, using for this the
recent continuum 1.2 mm data for NGC 4565 from Neininger et al. (1995). We
again choose the HSMN method for the centre and the gaussian fit for the warp
curve. The warp curve superimposed onto the contour levels of the image is shown
in figure (\ref{mm1})

\begin{figure}
\vspace{6cm}
\caption[]{Contour levels for NGC 4565 in 1.2 mm 
with its warp curve superimposed.}
\label{mm1}
\end{figure}

        In this figure we see that the right side of the galaxy is clearly
warped (as can also be seen from the calculated parameters). This effect is
also clear in 21 cm, but not so in the optical image (see Neininger et al.
(1995)), so this looks as if it there is a colour gradient within the warp.

        The warp parameters calculated for this case are:
        
\begin{eqnarray*}
w&=&0.00188 \\
w_r&=&0.00389 \\
w_l&=&0.000212 \\
\end{eqnarray*}

\section{Conclusions}

        We present here a new tool for the statistical study of geometrical 
properties of
warps. First we define new geometrical parameters which account for the size
and generic shape of the warp, and we then develope a software
which is able to calculate the warp curve and these parameters from
the image of a galaxy. We have shown several examples at different wavelengths 
in order to test the behaviour of both the software and the parameters. These
parameters are seen to be a useful tool in detecting some of the effects
predicted by the different theoretical models ( colour gradients, coherent
alignment of warps, etc...). This will allow to perform a
statistical analysis with a large number of galaxies, that could be a key
factor in understanding the warp phenomenon. For example, if we
limit such a study to spiral galaxies with inclination higher than 
$80^\circ$ (for which the method is certainly valid), a diameter
higher than $1.5 {\mathrm \; arcmin}$ and a total magnitude lower than 15
mag, the available sample contains 288 galaxies (according to the Lyon
Meudon Extraalactic Database), which is a sample large enough to allow
a statistical study.

        The small sample of galaxies considered here shows that the {\it warp
parameter} $w$ indeed represents the quantitative degree of deformation.
One of the galaxies known to be more warped, NGC 4013, has $w \simeq 0.023$.
Values much greater than this are not to be expected. It is expected that most warped galaxies
have a value of $w$ around 0.01.

        The value of the fitting parameters,and in particular the position at
which the warp begins and the asymptotic slope, are parameters whose mean value
and standard deviation could impose constraints to the different theoretical
models, and which inform us about the physical properties in regions external
to the disc, either about the dark halo or about the extragalactic medium.

\acknowledgements{
We would like to thank R. Bottema (Kapteyn Astronomical
Institute, Groningen), R. de Grijs (Kapteyn Astronomical Institute,
Groningen) and N. Neininger (MPI f\"ur Radioastronomie, Bonn) for
providing us the images of NGC 4013 in 21 cm , ESO 235-53 in the
optical, and NGC 4565 in 1.2 mm
respectively. We are also indebted to J. Cabrera (Rutgers University) for
valuable suggestions about the statistical treatment.}


\begin{thebibliography}{}

\bibitem{battaner}
Battaner E., 1995, in The Formation of the Milky Way, ed. E. J. Alfaro \& A. J.
Delgado. Cambridge University Press
\bibitem{binney}
Binney J., 1992, ARA\&A, 30, 51
\bibitem{bosma}
Bosma A., 1991, AJ, 86, 1791
\bibitem{bottema:1995}
Bottema R., 1995, A\&A, 295, 605
\bibitem{bottema:1996}
Bottema R., 1996, A\&A, 306, 345
\bibitem{briggs}
Briggs F. H., 1990, ApJ, 352, 15
\bibitem{christodoulou}
Christodoulou D. M., Tohline J. E., Steiman-Cameron T. Y., 1993, ApJ, 416, 74
\bibitem{combes}
Combes F., 1994, in The formation of Galaxies, ed. C. Mu\~noz-Tu\~n\'on \& F.
S\'anchez. Cambridge University Press.
\bibitem{degrijs}
de Grijs R. 1997, Ph. D. Thesis, Kapteyn Institute, Univ. of Groningen.
\bibitem{florido}
Florido E., Prieto M., Battaner E., Mediavilla E., S\'anchez-Saavedra M. L.,
1991, A\&A, 242, 301
\bibitem{neininger}
Neininger N. Gu\'elin M., Garc\'{\i}a-Burillo S., Zylka R., Wielebinski R.,
1995, A\&A
\bibitem{reshetnikov}
Reshetnikov V. P., 1995, Astron. Astrophys. Trans.
\bibitem{sanchez}
S\'anchez-Saavedra M. L., Battaner E., Florido E., 1990, MNRAS, 246, 458        
\end{thebibliography}
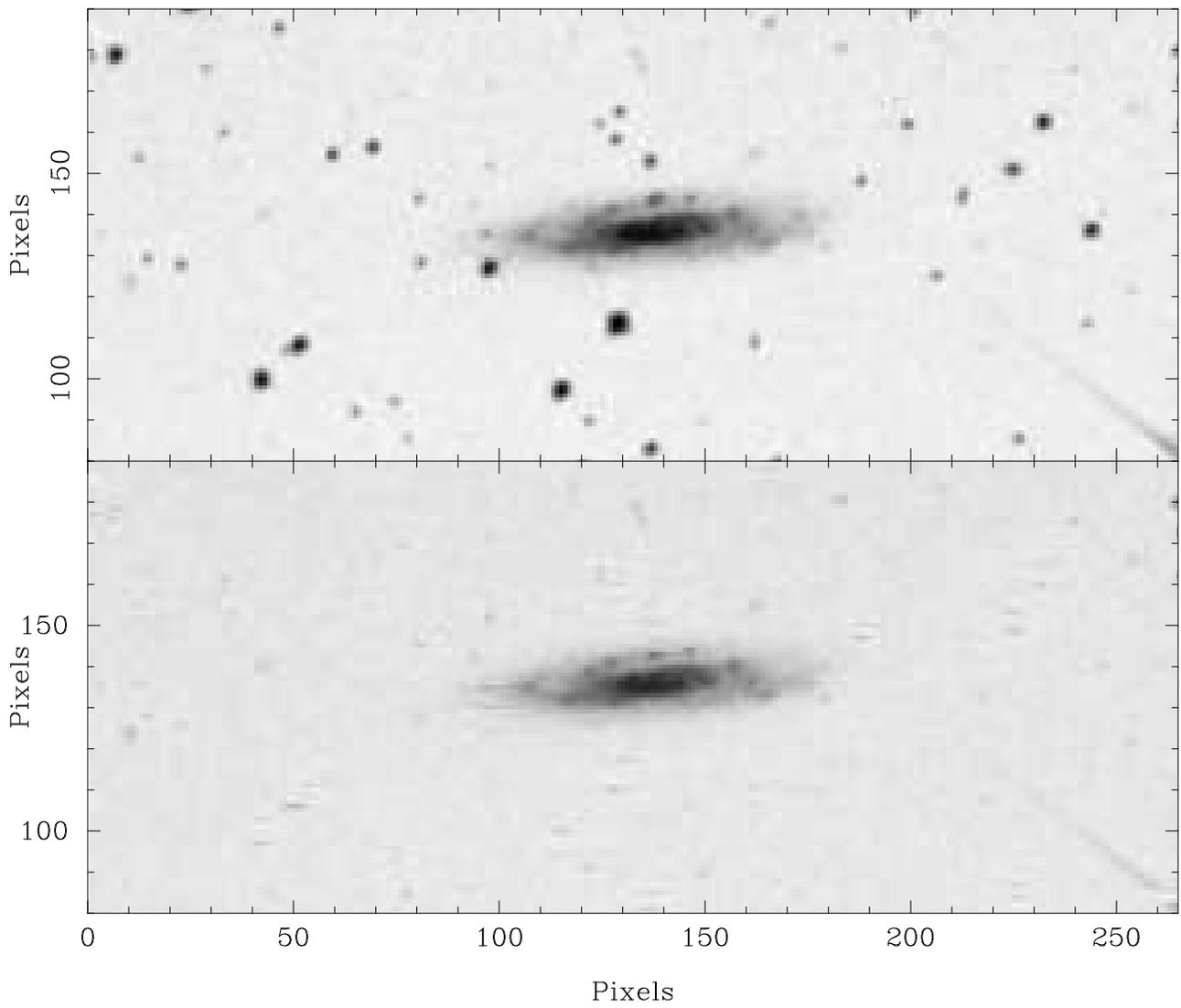
\end{document}